\documentclass[12pt]{article}
\begin{document}
\begin{center}
{\bf Deformed Quantum Field Theory, Thermodynamics at Low and High
Energies, and Gravity. I}\\ \vspace{5mm} A.E.Shalyt-Margolin
\footnote{E-mail: a.shalyt@mail.ru; alexm@hep.by}\\ \vspace{5mm}
\textit{National Center of Particles and High Energy Physics,
Bogdanovich Str. 153, Minsk 220040, Belarus}
\end{center}
PACS: 03.65, 05.20
\\
\noindent Keywords: quantum field theory with UV cutoff,
gravitational thermodynamics,deformed gravity \rm\normalsize
\vspace{0.5cm}
\begin{abstract}
The present work is a natural continuation of the previous paper arXiv: 0911.5597.
In this work, within the scope of the Generalized Uncertainty Principle, a model of the
high energy deformation  for a particular case
of Einstein's equations is developed.
In the process a thermodynamic description of General Relativity is used.
And the deformation is understood as an extension of a particular
theory by inclusion of one or several additional parameters in
such a way that the initial theory appears in the limiting  transition.
The possibility for the high energy deformation of
Einstein's equations within the scope of both equilibrium thermodynamics  and
non-equilibrium thermodynamics is examined.
\end{abstract}

\section{Introduction}
The present work is a natural continuation of the previous paper
arXiv: 0911.5597. It should be noted that there is a certain
discord between the modern development of quantum mechanics and
quantum field theory, on the one hand, and gravity, on the other
hand. In the last decade the researchers have come to an
understanding that in the process of studies into the physics of
the Early Universe (extremely high -- Planck's energies) the
fundamental physical theories, in particular quantum mechanics and
quantum field theory, should be changed. It is inevitable that in
these theories a fundamental length should be introduced. In so
doing the correspondence principle should be followed without
fail: at well-known low energies the theories involving the
fundamental length must present the conventional quantum mechanics
and quantum field theory with a high precision. The idea that a
quantum theory at the Planck scales must involve the fundamental
length has been put forward in the works devoted to a string
theory fairly a long time ago \cite{Ven1}. But since it is still
considered to be a tentative theory, some other indications have
been required. Fortunately, by the present time numerous
publications have suggested the appearance of the fundamental
length in the Early Universe with the use of various approaches
\cite{GUPg1}--\cite{Ahl1}. Of particular importance is the work
\cite{GUPg1}, where on the basis of a simple gedanken experiment
it is demonstrated that, with regard to the gravitational
interactions (Planck's scales)exhibited in the Early Universe
only, the Heisenberg Uncertainty Principle should be extended to
the Generalized Uncertainty Principle \cite{Ven1}--\cite{Ahl1}that
in turn is bound to bring forth the fundamental length on the
order of Planck's length. The advent of novel theories in physics
of the Early Universe is associated with the introduction of new
parameters, i.e. with a deformation of the well-known theories.
The deformation is understood as an extension of a particular
theory by inclusion of one or several additional parameters in
such a way that the initial theory appears in the limiting
transition \cite{Fadd}. Of course, in this case Heisenberg Algebra
is subjected to the corresponding deformation too. Such a
deformation may be based on the Generalized Uncertainty Principle
(GUP) \cite{Magg}--\cite{Kempf} as well as on the density matrix
deformation \cite{shalyt1}--\cite{shalyt9}.
\\ At the same time, the above-mentioned new deformation parameters
so far have not appeared in gravity despite the idea that they
should. The situation is that no evident efforts have been
undertaken to develop the high-energy (Planck's scale) gravity
deformations including the deformation parameters introduced in a
Quantum Theory of the Early Universe.
\\ In this paper, with   GUP held true, the possibility
for the high-energy gravity deformation is considered for a
specific case of Einstein's equations. As this takes place, the
parameter $\alpha$ appearing in the Quantum Field Theory (QFT)
with the UV cutoff (fundamental length) produced by the density
matrix deformation is used. There is no discrepancy of any kind as
the deformation parameter in the GUP-produced Heisenberg algebra
deformation is quite naturally expressed in terms of $\alpha$, and
this will be shown later (Section 2).  Besides, by its nature,
$\alpha$ is better applicable to study the high-energy deformation
of General Relativity because it is small, dimensionless (making
series expansion more natural), and the corresponding
representation îf Einstein's equations in its terms or its
deformation appear simple.
 Structurally, the paper is as follows.
In Sections 2 and 3 the approaches to the deformation of a quantum
theory at the Planck scales are briefly reviewed. In Section 4
together with various inferences a strategy is suggested to study
possible  high energy generalizations (deformations) of General
Relativity. Actually, Section 4 represents a short variant of
Section 5 given in  arXiv: 0911.5597 .   New results are presented
in Sections 5 and 6. A thermodynamic description of General
Relativity is used. The possibility for the high energy
deformation of Einstein's  equations is discussed within the scope
of both equilibrium thermodynamics and non-equilibrium
thermodynamics. In the latter case the approach is contemplated
only in terms of a nature of the cosmological constant.

\section{Quantum Theory at Planck's Scale}
In the last few years the researchers have come to the
understanding that studies of the Early Universe physics
(extremely high – Planck's energies) necessitate changes in the
fundamental physical theories, specifically quantum mechanics and
quantum field theory. Inevitably a fundamental length should be
involved in these theories \cite{Gar1}--\cite{Magg1}. This idea
has been first suggested by a string theory \cite{Ven1}. But it is
still considered to be a tentative theory without the experimental
status and merely an attractive model. However, the fundamental
length has been involved subsequently in more simple and natural
considerations \cite{GUPg1}.
\\The main approach to framing of Quantum Mechanics with fundamental
length (QMFL) and Quantum Field Theory with fundamental length
(QFTFL) (or with Ultraviolet (UV) cutoff) is that associated with
the Generalized Uncertainty Principle (GUP)
\cite{Ven1}--\cite{Kempf}:
\begin{equation}\label{GUP1}
\triangle x\geq\frac{\hbar}{\triangle p}+\alpha^{\prime}
l_{p}^2\frac{\triangle p}{\hbar}.
\end{equation}
 with the corresponding Heisenberg
algebra deformation produced by this principle
\cite{Magg}--\cite{Kempf}. \\Besides, in the works by the author
\cite{shalyt1}--\cite{shalyt10} an approach to the construction of
QMFL has been developed with the help of the deformed density
matrix, the density matrix deformation in QMFL being a starting
object called the density pro-matrix $\rho(\alpha)$ and
deformation parameter (additional parameter)
$\alpha=l_{min}^{2}/x^{2}$,$0<\alpha\leq1/4$ where $x$ is the
measuring scale and $l_{min}\sim l_{p}$
\cite{shalyt1},\cite{shalyt2}.
\\
\\The explicit form of the above-mentioned deformation gives an
exponential ansatz:
\begin{equation}\label{U26S}
\rho^{*}(\alpha)=exp(-\alpha)\sum_{i}\omega_{i}|i><i|,
\end{equation}
where all $\omega_{i}>0$ are independent of $\alpha$ and their sum
is equal to 1.
\\ In the corresponding deformed Quantum Theory (denoted as $QFT^{\alpha}$)
for average values we have
\begin{equation}\label{U26cS}
<B>_{\alpha}=exp(-\alpha)<B>,
\end{equation}
where $<B>$ - average in well-known QFT
\cite{shalyt6},\cite{shalyt7}.
 All the variables associated with the considered $\alpha$ -
deformed quantum field theory  are hereinafter marked with the
upper index $^{\alpha}$.
\\ Note that the deformation parameter
$\alpha$ is absolutely naturally represented as a ratio between
the squared UV and IR limits
\begin{equation}\label{U26dS}
\alpha=(\frac{UV}{IR})^{2},
\end{equation}
where UV is fixed and IR is varying.
\\It should be noted \cite{shalyt13} that in a series of the author's
works \cite{shalyt1}--\cite{shalyt10} a minimal
$\alpha$-deformation of QFT has been formed. By "minimal" it is
meant that no space-time noncommutativity was required, i.e. there
was no requirement for noncommutative operators associated with
different spatial coordinates
\begin{equation}\label{Concl1}
[X_{i},X_{j}]\neq 0, i\neq j.
\end{equation}
However, all the well-known deformations of QFT associated with
GUP (for example, \cite{Magg}--\cite{Kempf}) contain
(\ref{Concl1}) as an element of the corresponding deformed
Heisenberg algebra. Because of this, it is necessary to extend (or
modify) the above-mentioned minimal $\alpha$-deformation of QFT
--$QFT^{\alpha}$ \cite{shalyt1}--\cite{shalyt10} to some new
deformation $\widetilde{QFT}^{\alpha}$ compatible with GUP, as it
has been noted in \cite{shalyt13}. We can easily show that  QFT
parameter of deformations associated with GUP may be expressed in
terms of the parameter $\alpha$ that has been introduced in the
approach associated with the density matrix deformation. Here the
notation of \cite {Kim2} is used. Then
\begin{equation} \label{comm1}
[\vec{x}, \vec{p}]=i\hbar(1+\beta^2\vec{p}^2+...)
\end{equation}
and
\begin{equation}\label{comm2}
\Delta x_{\rm min}\approx
\hbar\sqrt{\beta}\sim l_{p}.
\end{equation}
Then from (\ref{comm1}),(\ref{comm2}) it follows that $\beta\sim
{\bf 1/p^{2}}$,  and for $x_{\rm min}\sim l_{p}$, $\beta$
corresponding to $x_{\rm min}$ is nothing else but
\begin{equation}\label{comm3}
\beta\sim  1/P_{pl}^{2},
\end{equation}
where $P_{pl}$ is Planck's momentum: $P_{pl}= \hbar/l_{p}$.
\\In this way $\beta$ is changing over the following interval:
\begin{equation}\label{comm4}
\lambda/P_{pl}^{2}\leq \beta<\infty,
\end{equation}
where $\lambda$  is  a numerical factor  and the second member in
(\ref{comm1}) is accurately reproduced in momentum representation
(up to the numerical factor) by $\alpha=l^{2}_{min}/l^{2}\sim
l^{2}_{p}/l^{2}=p^{2}/P_{pl}^{2}$
\begin{equation} \label{comm5}
[\vec{x},\vec{p}]=i\hbar(1+\beta^2\vec{p}^2+...)=i\hbar(1+a_{1}\alpha+a_{2}\alpha^{2}+...).
\end{equation}
\section{Some Inferences of Quantum Theories of the UV-cutoff}
The above-mentioned deformations of a quantum field theory at
Planck's scales have several important inferences. In particular,
a Quantum Field Theory (corresponding Heisenberg algebra
deformations) within the scope of GUP \cite{Magg}--\cite{Kempf}
suggests the high-energy quantum corrections for temperature and
entropy of the black holes \cite{acs}-- \cite{Nou}.In the recent
work \cite {Kim2} it has been demonstrated that the Holographic
Principle \cite{Hooft1}-- \cite{Bou3} is actually integrated in
the approach. Moreover, on the assumption that the cosmological
term $\Lambda$ is a dynamic quantity \cite{Ran1}-- \cite{Min4},
the Heisenberg Uncertainty principle derived in
\cite{Min1}--\cite{Min4} for the pair of conjugate variables
$(\Lambda,V)$:
\begin{equation}\label{CC1}
\Delta\Lambda\, \Delta V \sim \hbar,
\end{equation}
where $V$ is the space-time volume, may be extended up to GUP
\cite{shalyt-gup},\cite{shalyt-aip}. At least heuristically, this
result may account for a giant, by a factor of $\approx 10^{122}$,
discrepancy between the value of $\Lambda$ calculated within the
scope of conventional $QFT$ and the experimental value
\cite{Zel1},\cite{Wein1}.
\\ On the other hand, the parameter $\alpha$
and the corresponding deformation of a quantum field theory
$QFT^{\alpha}$ \cite{shalyt1}--\cite{shalyt10} also gives a good
explanation for this discrepancy
\cite{shalyt-aip},\cite{shalyt-vac},\cite{shalyt14}, at least
within the holographic principle and for the Holographic Dark
Energy Models (HDE) \cite{CKN} -- \cite{Odin}, as at the known
infrared limit (IR cut-off) of the Universe $L_{IR}\approx
10^{122}$ the quantity $\alpha$ is just equal to
$\alpha_{min}\approx 10^{122}$. Besides, an approach based on the
density matrix deformation suggests a phenomenological solution
for a number of problems in physics of black holes: Liouville
equation modification \cite{Hawk1}
(deformation)\cite{shalyt2}\cite{shalyt4},\cite{shalyt9},
information paradox of Hawking
\cite{Hawk2},\cite{shalyt4},\cite{shalyt5},\cite{shalyt9}, and
calculation of quantum corrections \cite{shalyt9}--\cite{shalyt13}
to a semiclassical Bekenstein-Hawking formula of the black hole
entropy \cite{Bek1}, \cite{Hawk3}.
 \\It should be noted that GUP (\ref{GUP1})
may be complemented by the Generalized Uncertainty Relation in
Thermodynamics (at Planck energy)
\cite{shalyt3},\cite{shalyt9},\cite{shalyt11}:
\begin{equation}\label{GUP2}
\Delta \frac{1}{T}\geq \frac{ k}{\Delta U}+\eta \left(\frac{\Delta
U}{E_{p}}\right)\, \frac{k_{B}}{E_{p}}+...
\end{equation}
where $T$ is the  ensemble temperature, $U$ is its internal
energy, $k_{B}$ is the Boltzmann constant, $E_{p}$ is the Planck
energy.In the recently published work \cite{Farmany})  the black
hole horizon temperature has been measured with the use of the
Gedanken experiment. In the process the Generalized Uncertainty
Relations in Thermodynamics (\ref{GUP2}) have been derived also.
Expression (\ref{GUP2}) has been considered in the monograph
\cite{Carroll} within the scope of the mathematical physics
methods.

\section{Gravitational Thermodynamics
in Low and High Energy and Deformed Quantum Theory} In the last
decade a number of very interesting works have been published. We
can primary name the works \cite{Padm1}--\cite{Padm12}, where
gravitation, at least for the â spaces with horizon, is directly
associated with thermodynamics and the results obtained
demonstrate a holographic character of gravitation. Of the
greatest significance is a pioneer work  \cite{Jac1}. For black
holes the association has been first revealed in
\cite{Bek1},\cite{Hawk1}, where related the black-hole event
horizon temperature to the surface gravitation. In \cite{Padm11},
has shown that this relation is not accidental and may be
generalized for the spaces with horizon.  As all the foregoing
results have been obtained in a semiclassical approximation, i.e.
for sufficiently low energies, the problem arises: how these
results are modified when going to higher energies. In the context
of this paper, the problem may be stated as follows: since we have
some infra-red (IR) cutoff $l_{max}$ and ultraviolet (UV) cutoff
$l_{min}$, we naturally have a problem how the above-mentioned
results on Gravitational Thermodynamics are changed for
\begin{equation}\label{GT1}
l \rightarrow l_{min}.
\end{equation}
According to Sections 2 and 3 of this paper, they should become
dependent on the deformation parameter $\alpha$. After all, in the
already mentioned in Section (formula (\ref{U26dS})) $\alpha$ is
indicated as nothing else but
\begin{equation}\label{GT2}
\alpha=\frac{l_{min}^{2}}{l^{2}}.
\end{equation}
In fact, in several papers \cite{acs}--\cite{Kim1} it has been
demonstrated that thermodynamics and statistical mechanics of
black holes in the presence of GUP (i.e. at high energies) should
be modified.  To illustrate, in \cite{Park} the Hawking
temperature modification has been computed in the asymptotically
flat space in this case in particular. It is easily seen that in
this case the deformation parameter $\alpha$ arises naturally.
Indeed, modification of the Hawking temperature is of the
following form(formula (10) in \cite{Park}):
\begin{equation}\label{GT3}
T_{GUP}=(\frac{d-3}{4\pi})\frac{\hbar r_{+}}{2\alpha^{\prime
2}l^{2}_{p}}[1-(1-\frac{4\alpha^{\prime 2}l_{p}^{2}}{
r_{+}^{2}})^{1/2}],
\end{equation}
where $d$ is the space-time dimension, and $r_+$ is the
uncertainty in the emitted particle position by the Hawking
effect, expressed as
\begin{equation}\label{GT4}
\Delta x_i \approx r_+
\end{equation}
and being nothing else but a radius of the event horizon;
$\alpha^{\prime}$ -- dimensionless constant from GUP. But as we
have $2\alpha^{\prime}l_{p}=l_{min}$, in terms of $\alpha$
 (\ref{GT3}) may be written in a natural way as follows:
\begin{equation}\label{GT5}
T_{GUP}=(\frac{d-3}{4\pi})\frac{\hbar \alpha^{-1}_{r_{+}}
}{\alpha^{\prime}l_{p}}[1-(1-\alpha_{r_{+}})^{1/2}],
\end{equation}
where $\alpha_{r_{+}}$- parameter $\alpha$ associated with the
IR-cutoff $r_{+}$. In such a manner $T_{GUP}$ is only dependent on
the constants including the fundamental ones and on the
deformation parameter $\alpha$.
\\The dependence of the black hole entropy on $\alpha$ may be derived
in a similar way. For a semiclassical approximation of the
Bekenstein-Hawking formula \cite{Bek1},\cite{Hawk1}
\begin{equation}\label{GT6}
S=\frac{1}{4}\frac{A}{l^{2}_{p}},
\end{equation}
where $A$ -- surface area of the event horizon, provided the
horizon event has radius $r_+$, then $A\sim r^{2}_+$ and
(\ref{GT6}) is clearly of the form
\begin{equation}\label{GT6.1}
S=\sigma \alpha^{-1}_{r_{+}},
\end{equation}
where $\sigma$ is some dimensionless denumerable factor. The
general formula for quantum corrections \cite{mv} given as
\begin{equation}\label{GT6.2}
S_{GUP} =\frac{A}{4l_{p}^{2}}-{\pi\alpha^{\prime 2}\over 4}\ln
\left(\frac{A}{4l_{p}^{2}}\right) +\sum_{n=1}^{\infty}c_{n}
\left({A\over 4 l_p^2} \right)^{-n}+ \rm{const}\;,
\end{equation}
where the expansion coefficients $c_n\propto \alpha^{\prime
2(n+1)}$ can always be computed to any desired order of accuracy
\cite{mv}, may be also written as a power series in
$\alpha^{-1}_{r_{+}}$   (or  Laurent series in $\alpha_{r_{+}}$)
\begin{equation}\label{GT6.3}
S_{GUP}=\sigma \alpha^{-1}_{r_{+}}-{\pi\alpha^{\prime 2}\over
4}\ln (\sigma \alpha^{-1}_{r_{+}}) +\sum_{n=1}^{\infty}(c_{n}
\sigma^{-n}) \alpha^{n}_{r_{+}}+ \rm{const}
\end{equation}
Note that here no consideration is given to the restrictions on
the IR-cutoff
\begin{equation}\label{GT7}
l\leq l_{max}
\end{equation}
and to those corresponding the extended uncertainty principle
(EUP) that leads to a minimal momentum \cite{Park}. This problem
will be considered separately in further publications of the
author.
\\A black hole is a specific example of the space with horizon.
It is clear that for other horizon spaces \cite{Padm11} a similar
relationship between their thermodynamics and  the deformation
parameter $\alpha$ should be exhibited.
\\Quite recently,
in a series of papers, and specifically in
\cite{Padm3}--\cite{Padm9}, it has been shown that Einstein
equations may be derived from the surface term of the GR
Lagrangian, in fact containing the same information as the bulk
term.
\\It should be noted that Einstein's
equations [at least for space with horizon] may be obtained from
the proportionality of the entropy and horizon area together with
the fundamental thermodynamic relation connecting heat, entropy,
and temperature \cite{Jac1}. In fact \cite{Padm3}-- \cite{Padm10},
this approach has been extended and complemented by the
demonstration of holographicity  for the gravitational action (see
also \cite{Padm11}).And in the case of Einstein-Hilbert gravity,
it is possible to interpret Einstein's equations as the
thermodynamic identity \cite{Padm12}:
\begin{equation}\label{GT8}
TdS = dE + PdV.
\end{equation}
The above-mentioned results in the last paragraph have been
obtained at low energies, i.e. in a semiclassical approximation.
Because of this, the problem arises how these results are changed
in the case of high energies? Or more precisely, how the results
of \cite{Jac1},\cite{Padm3}-- \cite{Padm12} are generalized in the
UV-limit?  It is obvious that, as in this case all the
thermodynamic characteristics become dependent on the deformation
parameter $\alpha$, all the corresponding results should be
modified (deformed) to meet the following requirements:
\\(a) to be clearly dependent on the deformation parameter
$\alpha$ at high energies;
\\
\\(b) to be duplicated, with high precision, at low energies
due to the suitable limiting transition;
\\
\\(c) Let us clear up what is meant by the adequate high energy $\alpha$-deformation
of Einstein's equations (General Relativity).
\\ The problem may be more specific.
\\ As, according to
\cite{Jac1},\cite{Padm11},\cite{Padm12} and some other works,
gravitation is greatly determined by thermodynamics and at high
energies the latter is a deformation of the classical
thermodynamics, it is interesting whether gravitation at high
energies (or what is the same, quantum gravity or Planck scale)is
being determined by the corresponding  deformed thermodynamics.
The formulae (\ref{GT5}) and (\ref{GT6.3}) are elements of the
high-energy $\alpha$-deformation in thermodynamics, a general
pattern of which still remains to be formed. Obviously, these
formulae should be involved in the general pattern giving better
insight into the quantum gravity, as they are applicable to black
mini-holes (Planck black holes) which may be a significant element
of such a pattern. But what about other elements of this pattern?
How can we generalize the results
\cite{Jac1},\cite{Padm11},\cite{Padm12}when the IR-cutoff tends to
the UV-cutoff (formula (\ref{GT1}))? What are modifications of the
thermodynamic identity (\ref{GT8}) in a high-energy deformed
thermodynamics and how is it applied in high-energy (quantum)
gravity? What are the aspects of using the Generalized Uncertainty
Relations in Thermodynamics
\cite{shalyt3},\cite{shalyt9},\cite{shalyt11} (\ref{GUP2})in this
respect? It is clear that these relations also form an element of
high-energy thermodynamics.
\\By authors opinion, the methods developed to solve the problem
of point (c) and elucidation of other above-mentioned problems may
form the basis for a new approach to solution of the quantum
gravity problem. And one of the keys to the {\bf quantum gravity}
problem is a better insight into the {\bf high-energy
thermodynamics}.
\section{$\alpha$--Representation of Einstein's Equations}
Let us consider $\alpha$-representation and high energy
$\alpha$-deformation of the Einstein's field equations for the
specific cases of horizon spaces (the point (c) of Section 4). In
so doing the results of the survey work (\cite{Padm13}
p.p.41,42)are used. Then, specifically, for a static, spherically
symmetric horizon in space-time described by the metric
\begin{equation}\label{GT9}
ds^2 = -f(r) c^2 dt^2 + f^{-1}(r) dr^2 + r^2 d\Omega^2
\end{equation}
the horizon location will be given by simple zero of the function
$f(r)$, at $r=a$.
\\  It is known that for horizon spaces one can introduce
the temperature that can be identified with an analytic
continuation to imaginary time. In the case under consideration
(\cite{Padm13}, eq.(116))
\begin{equation}\label{GT10}
k_BT=\frac{\hbar cf'(a)}{4\pi}.
\end{equation}
Therewith, the condition $f(a)=0$ and $f'(a)\ne 0$ must be
fulfilled.
\\ Then at the horizon $r=a$ Einstein's field equations
\begin{equation}\label{GT11}
\frac{c^4}{G}\left[\frac{1}{ 2} f'(a)a - \frac{1}{2}\right] = 4\pi
P a^2
\end{equation}
may be written as the thermodynamic identity
(\ref{GT8})(\cite{Padm13} formula (119))
\begin{equation}\label{GT12}
   \underbrace{\frac{{{\hbar}} cf'(a)}{4\pi}}_{\displaystyle{k_BT}}
    \ \underbrace{\frac{c^3}{G{{\hbar}}}d\left( \frac{1}{ 4} 4\pi a^2 \right)}_{
    \displaystyle{dS}}
  \ \underbrace{-\ \frac{1}{2}\frac{c^4 da}{G}}_{
    \displaystyle{-dE}}
 = \underbrace{P d \left( \frac{4\pi}{ 3}  a^3 \right)  }_{
    \displaystyle{P\, dV}}
\end{equation}
where $P = T^{r}_{r}$ is the trace of the momentum-energy tensor
and radial pressure. In the last equation $da$ arises in the
infinitesimal consideration of Einstein's equations when studying
two horizons distinguished by this infinitesimal quantity $a$ and
$a+da$ (\cite{Padm13} formula (118)).
\\ Now we consider(\ref{GT12})  in new notation expressing $a$
in terms of the corresponding deformation parameter $\alpha$. Then
we have
\begin{equation}\label{GT13}
a=l_{min}\alpha^{-1/2}.
\end{equation}
Therefore,
\begin{equation}\label{GT14}
f'(a)=-2l^{-1}_{min}\alpha^{3/2}f'(\alpha).
\end{equation}
Substituting this into (\ref{GT11}) or into (\ref{GT12}), we
obtain in the considered case of Einstein's equations in the
"$\alpha$--representation" the following:
\begin{equation}\label{GT16}
\frac{c^{4}}{G}(-\alpha f'(\alpha)-\frac{1}{2})=4\pi
P\alpha^{-1}l^{2}_{min}.
\end{equation}
Multiplying the left- and right-hand sides of the last equation by
$\alpha$, we get
\begin{equation}\label{GT16.1}
\frac{c^{4}}{G}(-\alpha^{2}f'(\alpha)-\frac{1}{2}\alpha)=4\pi
Pl^{2}_{min}.
\end{equation}
But since usually $l_{min}\sim l_{p}$ (that is just the case if
the Generalized Uncertainty Principle (GUP) is satisfied), we have
$l^{2}_{min}\sim l^{2}_{p}=G\hbar/c^{3}$. When selecting a system
of units, where $\hbar=c=1$, we arrive at $l_{min}\sim l_{p}=\surd
G$, and then (\ref{GT16}) is of the form
\begin{equation}\label{GT16.A}
-\alpha^{2}f'(\alpha)-\frac{1}{2}\alpha=4\pi P\vartheta^{2}G^{2},
\end{equation}
where $\vartheta=l_{min}/l_{p}$. L.h.s. of (\ref{GT16.A}) is
dependent on $\alpha$. Because of this, r.h.s. of (\ref{GT16.A})
must be dependent on $\alpha$ as well, i. e. $P=P(\alpha)$.
\begin{center}
{\bf Analysis of $\alpha$-Representation of Einstein's
Equations}
\end{center}

Now let us get back to (\ref{GT12}). In \cite{Padm13}
the low-energy case has been considered, for which (\cite{Padm13}
p.42 formula (120))
\begin{equation}\label{GT17.A}
 S=\frac{1}{ 4l_p^2} (4\pi a^2) = \frac{1}{ 4} \frac{A_H}{ l_p^2}; \quad E=\frac{c^4}{ 2G} a
    =\frac{c^4}{G}\left( \frac{A_H}{ 16 \pi}\right)^{1/2},
\end{equation}
where $A_H$ is the horizon area. In our notation (\ref{GT17.A})
may be rewritten as
\begin{equation}\label{GT17.A1}
 S= \frac{1}{4}\pi\alpha^{-1}; \quad E=\frac{c^4}{2G} a
 =\frac{c^4}{G}\left( \frac{A_H}{ 16 \pi}\right)^{1/2}=\frac{\vartheta}{2\surd G}\alpha^{1/2}.
\end{equation}
We proceed to two entirely different cases: low energy (LE) case
and high energy (HE) case. In our notation these are respectively
given by
\begin{center}
A)$\alpha\rightarrow 0$ (LE), B)$\alpha\rightarrow 1/4$ (HE),
\\C)$\alpha$ complies with the familiar scales and energies.
\end{center}
The case of C) is of no particular importance as it may be
considered within the scope of the conventional General
Relativity.
\\Indeed, in point A)$\alpha\rightarrow 0$ is not actually an exact
limit as a real scale of the Universe (Infrared (IR)-cutoff
$l_{max}\approx 10^{28}cm$), and then
\begin{center}
$\alpha_{min}\sim l_{p}^{2}/l^{2}_{max}\approx 10^{-122}$.
\end{center}
In this way A) is replaced by A1)$\alpha\rightarrow \alpha_{min}$.
In any case at low energies the second term in the left-hand side
(\ref{GT16.A}) may be neglected in the  infrared limit.
Consequently, at low energies (\ref{GT16.A}) is written as
\begin{equation}\label{GT16.LE}
-\alpha^{2}f'(\alpha)=4\pi P(\alpha)\vartheta^{2}G^{2}.
\end{equation}
Solution of the corresponding Einstein equation – finding of the
function $f(\alpha)=f[P(\alpha)]$ satisfying(\ref{GT16.LE}). In
this case formulae (\ref{GT17.A}) are valid as at low energies a
semiclassical approximation is true. But from (\ref{GT16.LE})it
follows that
\begin{equation}\label{GT16.solv}
f(\alpha)=-4\pi \vartheta^{2}G^{2}\int
\frac{P(\alpha)}{\alpha^{2}}d\alpha.
\end{equation}
On the contrary, knowing $f(\alpha)$, we can obtain
$P(\alpha)=T^{r}_{r}.$
\\ But it is noteworthy that, when studying the infrared modified gravity
\cite{Patil},\cite{Park1},\cite{Rub}, we have to make corrections for the considerations of point A1).

\section{Possible High Energy $\alpha$-Deformation of General Relativity}
Let us consider the high-energy case B). Here two variants are
possible.
\\
\\{\bf I. First variant}.
\\ In this case it is assumed that in the high-energy
(Ultraviolet (UV))limit the thermodynamic identity (\ref{GT12})(or
that is the same (\ref{GT8})is retained but now all the quantities
involved in this identity become $\alpha$-deformed. This means
that they appear in the $\alpha$-representation with quantum
corrections and are considered at high values of the parameter
$\alpha$, i.e. at $\alpha$ close to 1/4. In particular, the
temperature $T$ from equation (\ref{GT12}) is changed by $T_{GUP}$
(\ref{GT5}), the entropy $S$ from the same equation given by
semiclassical formula (\ref{GT17.A}) is changed by $S_{GUP}$
(\ref{GT6.3}), and so forth:
\begin{center}
$E\mapsto E_{GUP}, V\mapsto V_{GUP}$.
\end{center}
Then the high-energy $\alpha$-deformation of equation (\ref{GT12})
takes the form
\begin{equation}\label{GT8.GUP}
k_{B}T_{GUP}(\alpha)dS_{GUP}(\alpha)-dE_{GUP}(\alpha)=P(\alpha)dV_{GUP}(\alpha).
\end{equation}
Substituting into (\ref{GT8.GUP}) the corresponding quantities
\\$T_{GUP}(\alpha),S_{GUP}(\alpha),E_{GUP}(\alpha),V_{GUP}(\alpha),P(\alpha)$
and expanding them into a Laurent series in terms of $\alpha$,
close to high values of $\alpha$, specifically close to
$\alpha=1/4$, we can derive a solution for the high energy
$\alpha$-deformation of general relativity (\ref{GT8.GUP}) as a
function of $P(\alpha)$. As this takes place, provided at high
energies the generalization of (\ref{GT12}) to (\ref{GT8.GUP})is
possible, we can have the high-energy $\alpha$-deformation of the
metric. Actually, as from (\ref{GT12}) it follows that
\begin{equation}\label{GT8.GUP1}
f'(a)=\frac{4\pi k_{B}}{\hbar c}T=4\pi k_{B}T
\end{equation}
(considering that we have assumed $\hbar=c=1$), we get
\begin{equation}\label{GT8.GUP2}
f'_{GUP}(a)=4\pi k_{B}T_{GUP}(\alpha).
\end{equation}
L.h.s. of (\ref{GT8.GUP2}) is directly obtained in the
$\alpha$-representation. This means that, when $f'\sim T$, we have
$f'_{GUP}\sim T_{GUP}$ with the same factor of proportionality. In
this case the function $f_{GUP}$ determining the high-energy
$\alpha$-deformation of the spherically symmetric metric may be in
fact derived by the expansion of $T_{GUP}$, that is known from
(\ref{GT5}), into a Laurent series in terms of  $\alpha$ close to
high values of $\alpha$ (specifically close to $\alpha=1/4$), and
by the subsequent integration.
\\ It might be well to remark on the following.
\\
\\{\bf 6.1} As on going to high energies we use (GUP),
$\vartheta$ from equation (\ref{GT16.A})is expressed in terms of
$\alpha^{\prime}$--dimensionless constant from GUP
(\ref{GUP1}),(\ref{GT5}):$\vartheta=2\alpha^{\prime}.$
\\
\\{\bf 6.2} Of course, in all the formulae including $l_{p}$
this quantity must be changed by $G^{1/2}$ and hence $l_{min}$
by $\vartheta G^{1/2}=2\alpha^{\prime} G^{1/2}.$
\\
\\{\bf 6.3} As noted in the end of subsection 6.1,
and in this case also knowing all the high-energy deformed
quantities
$T_{GUP}(\alpha),S_{GUP}(\alpha),E_{GUP}(\alpha),V_{GUP}(\alpha)$,
we can find $P(\alpha)$ at $\alpha$ close to 1/4.
\\
\\{\bf 6.4} Here it is implicitly understood that the Ultraviolet
limit of Einstein's  equations is independent of the starting
horizon space. This assumption is quite reasonable. Because of
this, we use  the well-known formulae for the modification of
thermodynamics and statistical mechanics of black holes in the
presence of GUP \cite{acs}--\cite{Kim1}.
\\
\\{\bf 6.5} The use of the thermodynamic identity
(\ref{GT8.GUP}) for the description of the high energy deformation
in General Relativity implies that on going to the UV-limit of
Einstein's equations for horizon spaces in the thermodynamic
representation (consideration) we are trying to remain within the
scope of {\bf equilibrium statistical mechanics} \cite{Balesku1}
({\bf equilibrium thermodynamics}) \cite{Bazarov}. However, such
an assumption seems to be too strong. But some grounds to think so
may be found as well. Among other things, of interest is the
result from \cite{acs} that GUP may prevent black holes from their
total evaporation. In this case the Planck's remnants of black
holes will be stable, and when they are considered, in some
approximation the {\bf equilibrium thermodynamics} should be
valid. At the same time, by author's opinion these arguments are
rather weak to think that the quantum gravitational effects in
this context have been described only within the scope of {\bf
equilibrium thermodynamics}\cite{Bazarov}.
\\
\\{\bf II. Second variant}.
\\ According to the remark of {\bf 6.5},
it is assumed that the interpretation of Einstein's equations as a
thermodynamic identity (\ref{GT12}) is not retained on going to
high energies (UV--limit), i.e. at $\alpha\rightarrow 1/4$, and
the situation is adequately described exclusively by {\bf
non-equilibrium thermodynamics}\cite{Bazarov},\cite{Gyarm}.
Naturally, the question arises: which of the additional terms
introduced in (\ref{GT12}) at high energies may be leading to such
a description?
\\In the \cite{shalyt-gup},\cite{shalyt-aip} it has been shown that in case the cosmological
term $\Lambda$ is a dynamic quantity, it is small at low energies
and may be sufficiently large at high energies.
In the right-hand side of (\ref{GT16.A}) in the $\alpha$--representation the additional term
$GF(\Lambda(\alpha))$ is introduced:
 \begin{equation}\label{GT16.B}
 -\alpha^{2}f'(\alpha)-\frac{1}{2}\alpha=4\pi P\vartheta^{2}G^{2}-GF(\Lambda(\alpha)),
 \end{equation}
 where in terms of $F(\Lambda(\alpha))$ we denote the term including $\Lambda(\alpha)$ as a factor.
  Then its inclusion
in the low-energy case (\ref{GT11})(or in the $\alpha$
-representation (\ref{GT16.A})) has actually no effect on the
thermodynamic identity (\ref{GT12})validity, and consideration
within the scope of equilibrium thermodynamics still holds true.
It is well known that this is not the case at high energies as the
$\Lambda$-term may contribute significantly to make the "process"
non-equilibrium in the end \cite{Bazarov},\cite{Gyarm}.
\\ Is this the only cause for violation of the thermodynamic
identity (\ref{GT12}) as an interpretation of the high-energy
generalization of Einstein's equations? Further investigations are
required to answer this question.
\section{Conclusion}
This work presents the first steps to incorporation of the
deformation parameters of a quantum field theory at Planck's
scales into the high-energy deformation of General Relativity
(GR). Further, the corresponding calculations should follow with
an adequate interpretation. It is interesting to consider the high
energy $\alpha$-deformation of GR  in a more general case. The
problem is how far a thermodynamic interpretation of Einstein's
equations may be extended? We should remember that, as in all the
deformations considered a minimal length at the Planck level
$l_{min}\sim l_{p}$ has been involved, a minimal volume should
also be the case $V_{min}\sim V_{p}=l^{3}_{p}$, and this is of
particular importance for high energy thermodynamics (some
indications to this fact have been demonstrated in
\cite{shalyt-gup},\cite{shalyt-aip}).
\\Besides, in this paper we have treated QFT
with a minimal length, i.e. with the UV-cutoff. Consideration of
QFT with a minimal momentum (or IR-cutoff) \cite{Kim1}
necessitates an adequate extension of $\alpha$-deformation in QFT
with the introduction of new parameters significant in the
IR-limit.
\\It seems that some hints to a nature of such deformation may
be found from the works devoted to the infrared modification of
gravity \cite{Patil}--\cite{Rub}.

\end{document}